\documentclass[conference]{IEEEtran}
\IEEEoverridecommandlockouts

\usepackage{cite}
\usepackage[hidelinks,bookmarks=false]{hyperref}
\usepackage{amsmath,amssymb,amsfonts,bm}
\usepackage{booktabs}
\usepackage{array}
\usepackage{tabularx}
\usepackage{makecell}
\usepackage{graphicx}
\graphicspath{{figures/}}
\usepackage{textcomp}
\usepackage{xcolor}
\usepackage{url}
\usepackage{amsthm}
\usepackage{enumitem}
\usepackage{balance}
\usepackage{tikz}
\usetikzlibrary{positioning, arrows.meta, calc}

\theoremstyle{definition}

\newcommand{\VaR}{\operatorname{VaR}}
\newcommand{\calF}{\mathcal F}
\newcommand{\calC}{\mathcal C}
\newcommand{\ind}{\mathbf 1}

\makeatletter
\g@addto@macro\@maketitle{\vspace*{-2em}}
\makeatother
\setlist[itemize]{leftmargin=*, itemsep=1pt, topsep=2pt}
\setlist[enumerate]{leftmargin=*, itemsep=1pt, topsep=2pt}

\title{Marking-Aware Sequential VaR Recalibration for Standardized Option Books}

\author{\IEEEauthorblockN{Tenghan Zhong}
\IEEEauthorblockA{\textit{Department of Mathematics} \\
\textit{University of Southern California}\\
Los Angeles, CA, USA \\
tenghanz@usc.edu}
\and
\IEEEauthorblockN{Keyuan Wu}
\IEEEauthorblockA{\textit{Department of Mathematics} \\
\textit{University of Southern California}\\
Los Angeles, CA, USA \\
keyuanwu@usc.edu}}

\begin{document}
\maketitle

\begin{abstract}

Daily Value-at-Risk (VaR) for option books requires more than an accurate quantile forecast. It first requires a precise definition of the loss target. Before any model is evaluated, the protocol must fix the book construction rule, the marking rule for the next day, the loss scale, and the information set available at forecast time. Common pipelines instead apply VaR methods to underlying returns or preconstructed book loss series, leaving these operational choices outside the statistical target. We propose a marking-aware sequential VaR recalibration framework that targets normalized book-level loss directly, restricts the forecast state to information available at forecast time, and recalibrates an upper tail VaR using only past forecast residuals.

In out-of-sample evaluation on S\&P 500 index (SPX) and QQQ exchange-traded fund (ETF) options, the reference VaR undercovers all three books in both markets. Sequential VaR recalibration moves exceedance rates close to the target and delivers the best aggregate performance across books, with the lowest average violation, lowest pinball loss, and smallest maximum exceedance over rolling 50 trading day windows among the evaluated methods. Robustness checks preserve the same conclusion under strict direct marking, stricter book selection screens, and removal of the VaR floor. The result is also stable across alternative quantile learners, residual recalibration windows, and decay rates. These findings support marking-aware sequential VaR recalibration as a leakage-safe risk control layer for option-book VaR under realistic quote and marking frictions.

\end{abstract}

\begin{IEEEkeywords}
Value-at-Risk, option books, sequential VaR recalibration, model validation, derivatives risk, quote frictions.
\end{IEEEkeywords}

\bstctlcite{IEEEtran:BSTcontrol}
\section{Introduction}
\label{sec:intro}

Value-at-Risk (VaR) remains a standard market-risk measure \cite{bcbs1996backtesting}, but a daily VaR backtest is informative only when the tested loss corresponds to the position whose risk is being controlled. This requirement is especially important for option books. The controlled position is not a scalar index return. It is a multi-leg exposure whose value depends on the underlying price, the volatility surface, maturity, moneyness, liquidity, and contract-level quote availability.

A common response is to simplify the object before forecasting. One route applies classical VaR methods to the underlying return series or to a realized book loss series \cite{kupiec1995techniques,christoffersen1998evaluating,engle2004caviar,bollerslev1986garch}. Another route uses option-derived state variables such as implied volatility, variance risk premia, or liquidity summaries as predictors \cite{jiang2005model,carr2009variance,bollen2004net,garleanu2009demand}. These approaches are useful, but they do not by themselves define the book-level loss to be backtested. In a daily backtest, the contracts selected at date $t$ must be marked again at date $t+1$. Some contracts have no usable next-day quote, fail the liquidity screen, or need a fallback price because option quotes are contract-level and can be illiquid \cite{christoffersen2018illiquidity}. How these cases are handled changes the backtest: dropping the affected book-date observations changes the sample, while unreported fallback prices change the realized loss. Therefore, the next-day marking rule is part of the definition of the option-book VaR target.

Once the option-book loss target is fixed, the remaining question is how to calibrate the VaR forecast. Classical VaR backtests assess unconditional coverage, independence, and conditional coverage \cite{kupiec1995techniques,christoffersen1998evaluating}. Quantile regression and tree-based quantile models estimate conditional tail thresholds \cite{koenker1978regression,ke2017lightgbm}, while conformalized quantile regression and adaptive conformal inference motivate residual-based calibration under distributional drift \cite{romano2019cqr,gibbs2021adaptive,angelopoulos2021gentle}. These tools address calibration, but they do not by themselves connect the statistical target to the option-book implementation protocol. We address this combined problem by asking whether sequential VaR recalibration
can repair tail undercoverage for option books when the option-book VaR target is fixed.

In this paper, we evaluate one-day 10\% VaR for standardized SPX and QQQ option books using a model validation protocol that makes next-day marking explicit. The empirical target is the normalized next-day loss of three standardized books: an at-the-money straddle, a 25-delta risk reversal, and a 25-delta/10-delta short put spread. These books represent volatility, skew, and downside-spread exposures. A rolling Light Gradient Boosting Machine (LightGBM) quantile learner~\cite{ke2017lightgbm} produces the reference VaR threshold. We then add a one-sided recalibration layer that updates this threshold using only past forecast residuals. The evaluation asks whether sequential VaR recalibration restores near-target unconditional coverage and reduces violation severity for realized losses from these books without using the date-$t+1$ marks when issuing the date-$t$ forecast.

Our contributions are as follows:
\begin{itemize}[leftmargin=*,topsep=2pt,itemsep=1pt]
    \item \textbf{Marking-aware option-book VaR target.}
    We formulate VaR for standardized multi-leg option books under an explicit next-day marking hierarchy. The target fixes book construction, next-day valuation, loss normalization, and the forecast-time information set, rather than treating VaR as a quantile of an underlying return, a marginal option return, or an unspecified loss series.

    \item \textbf{Leakage-safe sequential VaR recalibration.}
    We embed a one-sided residual recalibration layer into the rolling
    option-book VaR protocol. The layer updates the reference VaR threshold using an empirical quantile of past forecast residuals, with no variable computed from the date-$t+1$ mark entering the date-$t$ forecast.

    \item \textbf{Dual-market validation under quote and marking frictions.}
    We validate the protocol on SPX index options and QQQ ETF options across three standardized book types. The evaluation benchmarks sequential recalibration against classical book-loss VaR baselines and tests whether the coverage repair is stable across marking, book construction, learner, and calibration choices. The diagnostics distinguish average coverage repair from remaining local exceedance clustering.
\end{itemize}

\section{Related Work}

\subsection{VaR Forecasting and Backtesting}

Classical VaR evaluation focuses on unconditional coverage, independence, and conditional coverage. Kupiec's likelihood-ratio test evaluates whether the empirical exceedance rate matches the target probability \cite{kupiec1995techniques}. Christoffersen's framework adds independence and conditional-coverage tests to detect exceedance clustering \cite{christoffersen1998evaluating}. Conditional Autoregressive Value-at-Risk (CAViaR) models the conditional quantile directly through a dynamic autoregressive quantile process \cite{engle2004caviar}. These tools are central to our evaluation, but their usual application is a scalar return series. We instead evaluate the next-day normalized loss of a standardized option book.

Generalized Autoregressive Conditional Heteroskedasticity (GARCH) VaR, historical VaR, and exponentially weighted moving average (EWMA) historical VaR remain important benchmarks because they are transparent and operationally common \cite{bollerslev1986garch,riskmetrics1996technical}. We therefore use historical VaR, EWMA historical VaR, CAViaR, and Student-$t$ GARCH VaR as book-loss baselines. The evaluation asks whether sequential VaR recalibration adds value beyond unconditional coverage by reducing violation severity and local undercoverage without relying on leaked information.

\subsection{Option-Book Risk and Marking Frictions}

Option prices encode forward-looking information about volatility, variance risk premia, and tail-risk compensation \cite{jiang2005model,carr2009variance,bollerslev2011tails}. At the same time, option books contain contract-level frictions. Net buying pressure can affect implied-volatility shapes, and demand-based option pricing links observed option prices to intermediary constraints and hedging pressure \cite{bollen2004net,garleanu2009demand}. For a daily VaR backtest, these frictions matter because strike discreteness, bid-ask spreads, and liquidity screens determine whether a contract can be marked on the next trading day \cite{christoffersen2018illiquidity}.

The resulting VaR target is a joint book loss distribution, not a marginal option-return distribution. A risk reversal and a short put spread can have similar headline deltas but very different tail behavior once book weights, hedge legs, and marks at $t+1$ are included. This motivates treating the book construction rule and the valuation rule at $t+1$ as part of the VaR target, rather than as implementation details outside the statistical object.

\subsection{Quantile Learning and Sequential Calibration}

Regression quantiles provide a direct conditional VaR target \cite{koenker1978regression}. Tree-based gradient boosting can capture nonlinear interactions in tabular financial state variables \cite{friedman2001greedy,chen2016xgboost,ke2017lightgbm}, which
motivates our choice of LightGBM as the reference learner. Conformalized quantile regression and adaptive conformal inference motivate residual-based coverage correction \cite{romano2019cqr,gibbs2021adaptive}. We adapt this logic to one-sided VaR thresholds. The layer leaves the reference learner unchanged, updates only from already realized forecast residuals, and can be applied to any rolling quantile forecast.

Recent work also connects conformal calibration with VaR-style marginal and conditional coverage testing, and with sequential portfolio VaR control under nonstationarity \cite{retzlaff2025coverage,schmitt2026portfolio}. Our setting differs because the risk object is the normalized next-day book loss, not an already observed return or portfolio loss series.

\section{Problem Formulation}
\label{sec:problem}

Let $t$ index trading dates and let $\calF_t$ be the close-of-day information set. For book type $b$, a deterministic selection rule maps the option chain $\calC_t$ into a book
\begin{equation}
B_t^{(b)}=\{(w_{t,\ell}^{(b)},a_{t,\ell}^{(b)}):\ell=1,\ldots,L_t^{(b)}\},
\end{equation}
where $L_t^{(b)}$ is the number of legs, $a_{t,\ell}^{(b)}$ is an option or hedge instrument, and $w_{t,\ell}^{(b)}$ is its weight. Let $M_t(\cdot)$ denote the date-$t$ mark, and let $\widetilde M_{t+1}(\cdot)$ denote the mark supplied by the next-day marking hierarchy for the old date-$t$ book. The date-$t$ value and implemented next-day value are
\begin{equation}
\begin{aligned}
V_t^{(b)}
&=\sum_{\ell=1}^{L_t^{(b)}} w_{t,\ell}^{(b)}M_t(a_{t,\ell}^{(b)}),\\
\widetilde V_{t+1}^{(b)}
&=\sum_{\ell=1}^{L_t^{(b)}}w_{t,\ell}^{(b)}
\widetilde M_{t+1}(a_{t,\ell}^{(b)}).
\end{aligned}
\end{equation}

Let $N_t^{(b)}>0$ denote the loss normalizer, assumed $\calF_t$-measurable and fixed before the next-day mark is observed. The realized normalized loss is
\begin{equation}
Y_{t+1}^{(b)}=\frac{V_t^{(b)}-\widetilde V_{t+1}^{(b)}}{N_t^{(b)}}.
\label{eq:normalized_loss}
\end{equation}
Positive values are losses. In the main specification, $N_t^{(b)}$ is the gross option premium of the option legs, the sum of absolute date-$t$ option premia excluding the spot hedge. Gross rather than net premium is used because the offsetting legs of the asymmetric books can leave a net premium near zero or of either sign, and this convention measures loss per unit of option premium rather than per full spot-hedged notional, which changes the interpretation of magnitudes.

The one-day VaR target is a threshold $q_{t,\alpha}^{(b)}$ satisfying
\begin{equation}
\Pr\left(Y_{t+1}^{(b)}>q_{t,\alpha}^{(b)}\mid \calF_t\right)\le \alpha,
\label{eq:var_target}
\end{equation}
with $\alpha=0.10$ in the main empirical study, which gives enough book-level
exceedances for stable finite-panel validation; see Section~\ref{subsec:alpha_robustness} for an $\alpha=0.05$ extension. This target is a book-level tail object. Contract-level marginal quantiles do not identify it because the loss depends on the joint marked distribution of all legs and hedge instruments.

To isolate the effect of marking, fix a book $b$ and date $t$. Write
$L=L_t^{(b)}$, $N=N_t^{(b)}$, $V_t=V_t^{(b)}$, and
$\widetilde V_{t+1}=\widetilde V_{t+1}^{(b)}$. Also write
$w_\ell=w_{t,\ell}^{(b)}$ and $a_\ell=a_{t,\ell}^{(b)}$. Let
$M_{t+1}^{\star}(\cdot)$ denote the benchmark exact next-day mark. Define
\[
V_{t+1}^{\star}
=
\sum_{\ell=1}^{L}w_\ell M_{t+1}^{\star}(a_\ell).
\]
The direct-mark and implemented normalized losses are
\[
Y^\star=\frac{V_t-V_{t+1}^{\star}}{N},
\qquad
Y=\frac{V_t-\widetilde V_{t+1}}{N}.
\]

\noindent\textbf{Marking distortion.}
Define
\begin{equation}
D^M
=
-\frac{1}{N}
\sum_{\ell=1}^{L}
w_\ell
\{\widetilde M_{t+1}(a_\ell)-M_{t+1}^{\star}(a_\ell)\}.
\label{eq:marking_distortion}
\end{equation}
Then $Y=Y^\star+D^M$. If
$|\widetilde M_{t+1}(a_\ell)-M_{t+1}^{\star}(a_\ell)|\le\varepsilon_\ell$
for all $\ell$, then
\[
|D^M|
\le
\frac{1}{N}
\sum_{\ell=1}^{L}|w_\ell|\varepsilon_\ell.
\]
If $\Pr(D^M\ge0\mid\calF_t)=1$, then every $\calF_t$-measurable threshold
$q_t$ satisfies
\[
\Pr(Y>q_t\mid\calF_t)
\ge
\Pr(Y^\star>q_t\mid\calF_t).
\]

The identity is by construction, and the bound follows from the triangle
inequality. Under $D^M\ge0$,
$\{Y^\star>q_t\}\subseteq\{Y>q_t\}$ for the same threshold; the inequality is
strict if
$\Pr(Y^\star\le q_t<Y^\star+D^M\mid\calF_t)>0$. Without a sign restriction, marking can move the implemented loss in either
direction. The main robust-marking sample uses direct next-day contract marks
when available and interpolation or nearest-neighbor fallback otherwise. We
therefore report direct-mark retention, the fraction of robust-marking book
dates with direct marks for every option leg; proxy-mark share, the share of
option-leg marks obtained through interpolation or fallback;
and a strict marking check that excludes any book date with an approximate
option mark.

\section{Method}
\label{sec:method}

For each book $b$, let $X_t^{(b)}$ denote the forecast-time covariate vector
and let $\widehat f_{b,t}$ denote the rolling conditional $(1-\alpha)$-quantile model
fitted before the date-$t$ forecast. The raw upper tail forecast is
$\widehat q_t^{\mathrm{raw},(b)}=\widehat f_{b,t}(X_t^{(b)})$. The floor-applied reference threshold is
\begin{equation}
\widehat q_t^{\mathrm{ref},(b)}
=
\max\{\widehat q_t^{\mathrm{raw},(b)},0\}.
\label{eq:ref_threshold}
\end{equation}
The same nonnegative floor is applied to the reference and residual-adjusted thresholds to avoid a negative desk VaR limit. We treat this as an implementation choice and report no-floor robustness.

After observing $Y_{s+1}^{(b)}$, define the residual as $R_s^{(b)}=Y_{s+1}^{(b)}-\widehat q_s^{\mathrm{ref},(b)}$. At date $t$, let $\mathcal I_t$ be the recent residual window and let $\eta\ge0$ be the time-decay rate. In the main specification, $\mathcal I_t$ contains at most the most recent 126 realized residuals, the weighted-window adjustment is used once at least 30 residuals are available, and $\eta=0.01$. For each past forecast date $s\in\mathcal I_t$, define the exponentially decaying weight
\begin{equation}
\omega_{t,s}=\frac{\exp\{-\eta(t-s)\}}{\sum_{u\in\mathcal I_t}\exp\{-\eta(t-u)\}}.
\end{equation}
For each book $b$, define the weighted empirical residual cumulative distribution function (CDF) by
$\widehat G_t^{\omega,(b)}(x)=\sum_{s\in\mathcal I_t}\omega_{t,s}\ind\{R_s^{(b)}\le x\}$.
We use the weighted quantile defined by the generalized inverse
$Q_\tau^\omega(\{R_s^{(b)}:s\in\mathcal I_t\})=\inf\{x:\widehat G_t^{\omega,(b)}(x)\ge \tau\}$.
The residual quantile adjustment and recalibrated threshold are
\begin{equation}
\begin{aligned}
\widehat\Delta_t^{(b)}
&= Q_{1-\alpha}^{\omega}\left(\{R_s^{(b)}:s\in \mathcal I_t\}\right),\\
\widehat q_t^{\mathrm{recal},(b)}
&= \max\{\widehat q_t^{\mathrm{ref},(b)}+\widehat\Delta_t^{(b)},0\}.
\end{aligned}
\label{eq:recal_threshold}
\end{equation}
During the initial warm-up period, if the weighted quantile cannot be estimated, we use the unweighted residual quantile. Section~\ref{sec:results} reports calibration memory ablations that vary the residual-window length and exponential decay rate while holding the reference forecasts fixed.

\noindent\textbf{Residual drift interpretation.}
The following interpretation applies on dates for which
$\widehat\Delta_t=Q_{1-\alpha}^{\omega}$ is computed from the current residual
window. Fix a book $b$ and suppress superscripts. Let $G_t$ be the conditional CDF of
the next reference residual
$R_t=Y_{t+1}-\widehat q_t^{\mathrm{ref}}$, and let
$\widehat G_t^\omega$ be the weighted empirical residual CDF induced by the
same residual window and weights used in $Q_{1-\alpha}^{\omega}$. Define
\[
d_t=\sup_x\left|G_t(x)-\widehat G_t^\omega(x)\right|.
\]
Then
\[
\Pr\{Y_{t+1}>\widehat q_t^{\mathrm{recal}}\mid\calF_t\}
\le \alpha+d_t.
\]
Indeed, by the generalized inverse weighted quantile definition,
$\widehat G_t^\omega(\widehat\Delta_t)\ge 1-\alpha$, so
$G_t(\widehat\Delta_t)\ge 1-\alpha-d_t$. Hence
$\Pr(R_t>\widehat\Delta_t\mid\calF_t)\le \alpha+d_t$.
Since
$\widehat q_t^{\mathrm{recal}}
=\max\{\widehat q_t^{\mathrm{ref}}+\widehat\Delta_t,0\}
\ge \widehat q_t^{\mathrm{ref}}+\widehat\Delta_t$,
we have
\[
\{Y_{t+1}>\widehat q_t^{\mathrm{recal}}\}
\subseteq
\{R_t>\widehat\Delta_t\}.
\]
Thus the same $\alpha+d_t$ bound applies. Since $\alpha$ is fixed and only
the empirical residual quantile adjustment is updated, this bound diagnoses
drift in the residual distribution rather than providing an adaptive conformal
coverage guarantee.

\noindent\textbf{Operational protocol.}
Figure~\ref{fig:operational_protocol} summarizes the following procedure
for each prediction date and book:
\begin{enumerate}
    \item build the standardized book at date $t$ from the available option chain;
    \item compute $X_t^{(b)}$ from current market variables, current book
    descriptors, lagged book loss summaries, and lagged marking-feasibility
    summaries;
    \item estimate or update the rolling $(1-\alpha)$ conditional quantile model using
    the most recent 252 valid training observations;
    \item apply the residual quantile adjustment in (\ref{eq:recal_threshold});
    \item after $t+1$, mark the book selected at date $t$, compute the realized
    residual, and add it to the residual history used by future forecasts.
\end{enumerate}

\begin{figure}[!t]
\centering
\resizebox{0.92\columnwidth}{!}{%
\begin{tikzpicture}[
    >=Stealth,
    font=\scriptsize,
    every node/.style={align=center},
    box/.style={
        rectangle,
        rounded corners=2.4pt,
        draw=black!65,
        thick,
        inner xsep=4pt,
        inner ysep=3.4pt,
        fill=white
    },
    stage/.style={
        box,
        draw=black!45,
        fill=black!3,
        font=\bfseries\scriptsize,
        text width=3.20cm,
        minimum height=6.0mm
    },
    fbox/.style={
        box,
        text width=3.20cm,
        minimum height=9.2mm
    },
    statebox/.style={
        fbox,
        draw=orange!70!black,
        fill=orange!6
    },
    calbox/.style={
        fbox,
        draw=purple!75!black,
        fill=purple!8,
        very thick
    },
    ubox/.style={
        fbox,
        draw=green!50!black,
        fill=green!6
    },
    futurebox/.style={
        fbox,
        draw=black!55,
        fill=black!3,
        dashed
    },
    arr/.style={
        -{Stealth[length=1.45mm,width=.95mm]},
        thick,
        draw=black!72,
        shorten <=1.3pt,
        shorten >=1.3pt
    },
    softarr/.style={
        -{Stealth[length=1.25mm,width=.85mm]},
        semithick,
        draw=black!50,
        dashed,
        shorten <=1.3pt,
        shorten >=1.3pt
    }
]

\def\xsep{4.05cm}
\def\ysep{1.12cm}

\node[stage] (hL) at (0,0) {Forecast at $t$};
\node[stage] (hR) at (\xsep,0) {Observed at $t+1$};
\draw[arr] (hL.east) -- (hR.west);

\node[fbox] (book) at (0,-\ysep) {%
\textbf{Book $B_t^{(b)}$}\\[-.2ex]
standardized exposure
};

\node[statebox] (state) at (0,-2*\ysep) {%
\textbf{State $X_t^{(b)}$}\\[-.2ex]
ex-ante variables $+$ lags
};

\node[fbox] (ref) at (0,-3*\ysep) {%
\textbf{Reference VaR}\\[-.2ex]
rolling $(1-\alpha)$ quantile
};

\node[calbox] (cal) at (0,-4*\ysep) {%
\textbf{Recalibrated VaR}\\[-.2ex]
ref. VaR $+$ residual adj.
};

\node[ubox] (mark) at (\xsep,-\ysep) {%
\textbf{Next-day mark}\\[-.2ex]
old-book value $\widetilde V_{t+1}^{(b)}$
};

\node[ubox] (loss) at (\xsep,-2*\ysep) {%
\textbf{Loss $Y_{t+1}^{(b)}$}\\[0.15ex]
normalized realized loss
};

\node[ubox] (resid) at (\xsep,-3*\ysep) {%
\textbf{Residual $R_t^{(b)}$}\\[-.2ex]
realized loss $-$ ref. VaR
};

\node[futurebox] (future) at (\xsep,-4*\ysep) {%
\textbf{Residual history update}\\[-.2ex]
after observation only
};

\draw[arr] (book.south) -- (state.north);
\draw[arr] (state.south) -- (ref.north);
\draw[arr] (ref.south) -- (cal.north);

\draw[arr] (mark.south) -- (loss.north);
\draw[arr] (loss.south) -- (resid.north);
\draw[arr] (resid.south) -- (future.north);

\draw[softarr] (book.east) -- (mark.west);

\end{tikzpicture}%
}
\caption{Operational timing for VaR on option books with explicit marking. Forecasts at date $t$ use only variables available at forecast time and lagged summaries. Marks and realized residuals are observed at $t+1$ and only affect future residual updates.}
\label{fig:operational_protocol}
\end{figure}

Table~\ref{tab:data_summary} summarizes the screened option-chain scale and the forecast-time information restriction. The vector $X_t^{(b)}$ contains date-$t$ market and book descriptors plus lagged summaries only. It contains no variable computed from $\widetilde M_{t+1}$, $Y_{t+1}^{(b)}$, or the date-$t+1$ fallback marking decision for the date-$t$ book.

\section{Experimental Design}
\label{sec:experiment}

We use daily SPX index option chains and QQQ ETF option chains from 2018 through August 2025, sourced from OptionMetrics IvyDB US. Each option chain is merged with the underlying spot series, zero-coupon yields, dividend yields, the Cboe Volatility Index (VIX), and the Cboe Three-Month Volatility Index (VIX3M, historically distributed in some data feeds as VXV) when available. We retain contracts with 14 to 120 calendar days to expiry, positive bid, ask
above bid, midpoint above 0.05, positive implied volatility, relative bid-ask
spread no larger than 0.50, and positive open interest or volume. Let $K$ denote the strike and $F$ the date-$t$ forward price. SPX uses $\log(K/F)\in[-0.20,0.10]$, while QQQ uses
$\log(K/F)\in[-0.35,0.25]$ so that the same quote screens retain enough
25-delta and 10-delta ETF contracts for book construction.

The three books are nearest-feasible 30-day proxies: an at-the-money (ATM)
straddle, a 25-delta risk reversal (RR), and a 25-delta/10-delta short put
spread. In all tables, 25d and 10d denote 25-delta and 10-delta option legs. The target expiry is the listed maturity nearest to 30 calendar days.
The risk reversal is long a 25-delta call and short a 25-delta put,
with a spot hedge for residual delta. The put spread is short a 25-delta put and long a 10-delta put, also delta-hedged. The main experiment uses nearest-feasible construction because option chains
thin unevenly through time. To measure book-identity risk, we also define a
strict target-quality screen. For ATM legs, the screen requires
$|\log(K/F)|\le 0.05$. For delta-targeted legs, the selected leg's date-$t$
delta must lie within $\pm 0.10$ of the target, and maturity error must be no
larger than seven calendar days. We call the robustness run that imposes this
screen the strict quality check.

For next-day marks, the robust specification follows a fixed hierarchy. It first uses the same option identifier, or the same underlying, option type, expiry, and strike, when that contract is quoted at date $t+1$. If no direct quote is available, it falls back to same-expiry interpolation across nearby strikes and then to the nearest available option mark within a seven-calendar-day maturity gap. The strict marking robustness check keeps only observations with direct next-day contract marks for all option legs. Spot hedge legs are marked using the next-day underlying close.

\noindent\textbf{Reference learner choice.}
We use LightGBM quantile regression~\cite{ke2017lightgbm} as the reference learner. The rolling window has 252 observations, the model is re-estimated every five days, and the feature set contains 79 mixed scale tabular variables spanning market state, book descriptors, lagged book-loss summaries, and lagged marking-feasibility summaries. Tree boosting is a natural choice for this heterogeneous tabular
setting relative to higher-capacity deep tabular
models~\cite{shwartz2022tabular}, and LightGBM provides a quantile
objective with an $\alpha$ parameter~\cite{lightgbm_docs_quantile}.
Section~\ref{sec:results} reports robustness checks with gradient-boosted
regression trees (GBRT)~\cite{friedman2001greedy} and XGBoost~\cite{chen2016xgboost} as alternative boosted-tree
quantile learners.

Baselines are rolling historical VaR, EWMA historical VaR with decay $\lambda=0.97$, CAViaR with the symmetric absolute-value dynamic quantile specification~\cite{engle2004caviar}, and GARCH-$t$ VaR~\cite{bollerslev1986garch,bollerslev1987t}. We evaluate empirical exceedance, average violation, upper-tail pinball loss, rolling 50-day exceedance, and Kupiec-Christoffersen backtests~\cite{kupiec1995techniques,christoffersen1998evaluating}. For a forecast $\widehat q_t$, exceedance is $\ind\{Y_{t+1}>\widehat q_t\}$, where $\ind\{\cdot\}$ denotes the indicator function, and average violation is the sample mean of $(Y_{t+1}-\widehat q_t)_+$, where $x_+=\max\{x,0\}$. Pinball loss is the upper-tail quantile check loss at $\tau=1-\alpha$, with $\tau=0.90$ in the main 10\% study \cite{koenker1978regression,gneiting2011making}.

\begin{table}[t]
\caption{Option-chain sample and forecast-time variables. Raw and clean observations are contract-date option rows before and after quote screens. Retained share is clean/raw. Variables are counted when the date-$t$ VaR forecast is issued.}
\label{tab:data_summary}
\centering
\footnotesize
\setlength{\tabcolsep}{5.0pt}
\renewcommand{\arraystretch}{1.03}
\begin{tabular}{@{}lrr@{}}
\toprule
Panel item & \multicolumn{1}{c}{SPX} & \multicolumn{1}{c}{QQQ} \\
\midrule
Raw observations   & 35,968,650 & 11,614,397 \\
Clean observations & 4,363,137  & 1,797,717  \\
Retained share     & 0.121      & 0.155      \\
\midrule
Variables in $X_t^{(b)}$  & 79 & 79 \\
Date-$t+1$ mark variables & 0  & 0  \\
\bottomrule
\end{tabular}
\end{table}

\begin{table*}[t]
\caption{Book sample and marking diagnostics. Direct-mark retention and
proxy-mark share are defined in Section~\ref{sec:problem}.}
\label{tab:book_quality_marking}
\centering
\footnotesize
\setlength{\tabcolsep}{3.8pt}
\begin{tabular*}{\textwidth}{@{\extracolsep{\fill}}llccccc@{}}
\toprule
Market & Book & Main $n$ & \makecell{Target-quality\\share} & \makecell{Target-quality\\pass $n$} & \makecell{Direct-mark\\retention} & \makecell{Proxy-mark\\share} \\
\midrule
SPX & ATM straddle & 1,077 & 0.710 & 765 & 0.775 & 0.155 \\
SPX & 25d risk reversal & 1,106 & 0.589 & 651 & 0.678 & 0.209 \\
SPX & 25d/10d put spread & 978 & 0.862 & 843 & 0.802 & 0.110 \\
QQQ & ATM straddle & 1,484 & 0.415 & 616 & 0.719 & 0.128 \\
QQQ & 25d risk reversal & 1,497 & 0.672 & 1,006 & 0.911 & 0.048 \\
QQQ & 25d/10d put spread & 988 & 0.970 & 958 & 0.972 & 0.016 \\
\bottomrule
\end{tabular*}
\end{table*}

\begin{table*}[htp]
\caption{Main out-of-sample VaR results. Here $n$ is the out-of-sample book-date count. Reference is the floor-applied LightGBM threshold in (\ref{eq:ref_threshold}); Recalibrated is the sequentially adjusted threshold in (\ref{eq:recal_threshold}). Pinball loss is the upper-tail quantile check loss at $\tau=0.90$. Max 50-day exceedance is the maximum rolling 50-trading-day empirical exceedance rate.}
\label{tab:main_results}
\centering
\footnotesize
\setlength{\tabcolsep}{3.6pt}
\begin{tabular*}{\textwidth}{@{\extracolsep{\fill}}llccccccccc@{}}
\toprule
Market & Book & $n$
& \multicolumn{2}{c}{Exceedance}
& \multicolumn{2}{c}{Average violation}
& \multicolumn{2}{c}{Pinball loss}
& \multicolumn{2}{c}{\makecell{Max 50-day\\exceedance}} \\
\cmidrule(lr){4-5}
\cmidrule(lr){6-7}
\cmidrule(lr){8-9}
\cmidrule(lr){10-11}
& & & Reference & Recalibrated & Reference & Recalibrated & Reference & Recalibrated & Reference & Recalibrated \\
\midrule
SPX & ATM straddle & 1,077 & 0.195 & 0.111 & 0.013 & 0.007 & 0.027 & 0.025 & 0.32 & 0.24 \\
SPX & 25d risk reversal & 1,106 & 0.170 & 0.099 & 0.046 & 0.040 & 0.085 & 0.085 & 0.30 & 0.22 \\
SPX & 25d/10d put spread & 978 & 0.147 & 0.108 & 0.005 & 0.003 & 0.010 & 0.010 & 0.36 & 0.26 \\
QQQ & ATM straddle & 1,484 & 0.221 & 0.106 & 0.014 & 0.005 & 0.026 & 0.022 & 0.42 & 0.22 \\
QQQ & 25d risk reversal & 1,497 & 0.176 & 0.108 & 0.006 & 0.004 & 0.011 & 0.010 & 0.32 & 0.20 \\
QQQ & 25d/10d put spread & 988 & 0.168 & 0.105 & 0.005 & 0.003 & 0.008 & 0.008 & 0.30 & 0.20 \\
\bottomrule
\end{tabular*}
\end{table*}

Table~\ref{tab:book_quality_marking} reports construction and marking
diagnostics. Main $n$ is the main robust-marking out-of-sample book date count. Target-quality share is the fraction of the main robust-marking
sample that passes the strict target-quality screen, and target-quality pass
$n$ is the corresponding number of passing book dates. The screen is most
binding for QQQ ATM straddles and SPX risk reversals, with pass shares of
0.415 and 0.589. Direct-mark retention ranges from 0.678 to 0.972, and
proxy-mark share ranges from 0.016 to 0.209.

\section{Results}
\label{sec:results}

\subsection{Quantile Learning and Coverage Repair}
\begin{table*}[htp]
\caption{Aggregate comparison with classical VaR baselines. Exceedance, average violation, pinball loss, and average VaR are pooled across book-date observations within each market. Max 50-day exceedance is the maximum over book-specific rolling 50-trading-day exceedance rates.}
\label{tab:baseline_results}
\centering
\footnotesize
\setlength{\tabcolsep}{4.8pt}
\begin{tabular*}{\textwidth}{@{\extracolsep{\fill}}llccccc@{}}
\toprule
Market & Method
& Exceedance
& \makecell{Average\\violation}
& \makecell{Pinball\\loss}
& \makecell{Average\\VaR}
& \makecell{Max 50-day\\exceedance} \\
\midrule
SPX & Historical VaR & 0.105 & 0.026 & 0.0464 & 0.089 & 0.42 \\
SPX & EWMA historical VaR & 0.103 & 0.020 & 0.0442 & 0.128 & 0.28 \\
SPX & CAViaR & 0.103 & 0.024 & 0.0455 & 0.095 & 0.34 \\
SPX & GARCH-$t$ VaR & 0.072 & 0.028 & 0.0525 & 0.130 & 0.42 \\
SPX & LightGBM reference VaR & 0.171 & 0.022 & 0.0420 & 0.083 & 0.36 \\
SPX & LightGBM recalibrated VaR & 0.106 & 0.017 & 0.0415 & 0.125 & 0.26 \\
\midrule
QQQ & Historical VaR & 0.111 & 0.007 & 0.0175 & 0.124 & 0.36 \\
QQQ & EWMA historical VaR & 0.104 & 0.005 & 0.0162 & 0.128 & 0.26 \\
QQQ & CAViaR & 0.109 & 0.006 & 0.0170 & 0.127 & 0.38 \\
QQQ & GARCH-$t$ VaR & 0.108 & 0.007 & 0.0176 & 0.122 & 0.36 \\
QQQ & LightGBM reference VaR & 0.191 & 0.009 & 0.0157 & 0.086 & 0.42 \\
QQQ & LightGBM recalibrated VaR & 0.106 & 0.004 & 0.0140 & 0.116 & 0.22 \\
\bottomrule
\end{tabular*}
\end{table*}

\begin{figure}[htp]
\centering
\includegraphics[width=\columnwidth]{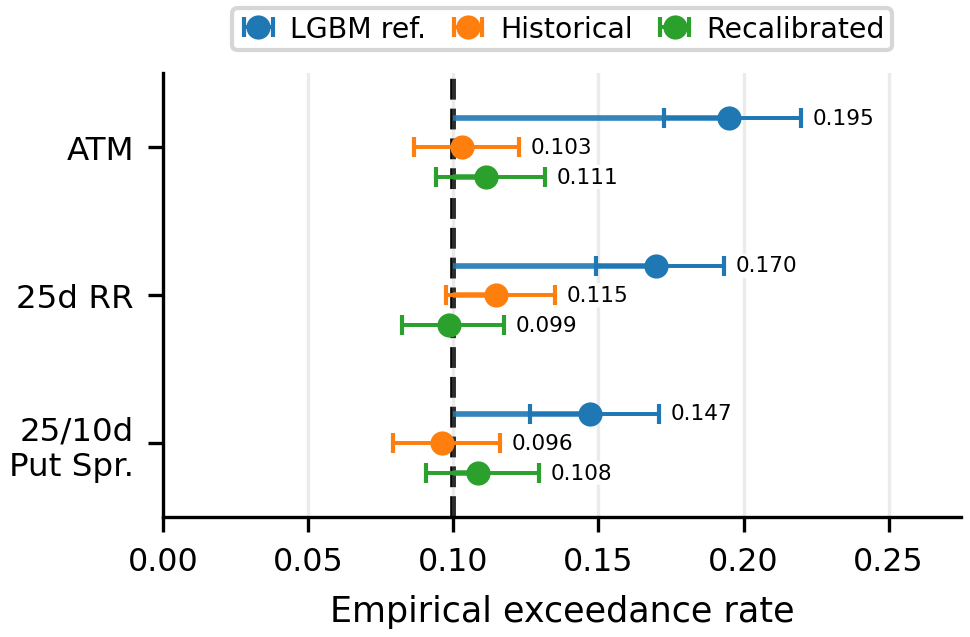}
\caption{SPX exceedance rates with 95\% binomial intervals.
Dashed line: 10\% target. Historical VaR is shown per book; aggregate
baselines are in Table~\ref{tab:baseline_results}. Reference VaR
undercovers all books; recalibration restores near-target coverage.}
\label{fig:spx_overall_ci}
\end{figure}

Table~\ref{tab:main_results} reports the central module ablation and
dual-market result. The reference and recalibrated forecasts use the same
LightGBM reference learner; the only added component is the residual
recalibration layer in (\ref{eq:recal_threshold}). Figure~\ref{fig:spx_overall_ci}
visualizes the SPX rows. The LightGBM reference VaR undercovers all books,
while sequential recalibration moves exceedance close to the 10\% target in
both markets.

The recalibrated VaR passes the Kupiec unconditional-coverage test for all six book-market pairs. Aggregating the three books within each market, the reference LightGBM forecasts are rejected by the unconditional-coverage test, with exceedance rates of 0.171 in SPX and 0.191 in QQQ. The remaining independence issue is discussed in Section~\ref{sec:validation_threats}.

\subsection{Comparison with Classical Book-Loss VaR Baselines}

Classical book-loss VaR baselines are strong competitors as they use recent realized book losses directly. Table~\ref{tab:baseline_results} compares these baselines with the LightGBM reference VaR and the recalibrated VaR after aggregating book-date observations across the three books within each market. The comparison shows why average coverage alone is not enough. Historical VaR, EWMA historical VaR, CAViaR, and GARCH-$t$ often deliver reasonable aggregate exceedance rates, especially in QQQ, but they do not dominate on violation severity or local stability. The main exception is SPX GARCH-$t$, whose aggregate exceedance rate is only 0.072, indicating a materially conservative threshold; this lower exceedance rate does not translate into better severity or pinball performance, since the method has the largest SPX average violation (0.028) and pinball loss (0.0525).

The LightGBM reference learner has useful conditional signal, with lower
pinball loss than historical VaR in both markets. However, it is not well
calibrated as a stand-alone VaR engine: aggregate exceedance rates are 0.171 in
SPX and 0.191 in QQQ, well above the 0.10 target. This undercoverage is
plausible in the rolling design because tail calibration at the 0.90 quantile
is informed by roughly 25 upper-tail observations in each 252-day window before
conditioning on book descriptors and market state. The recalibration layer repairs this level error
while retaining the useful part of the conditional forecast. On the reported
aggregate metrics, the recalibrated VaR attains the lowest average violation,
lowest pinball loss, and smallest maximum rolling 50-day exceedance among the
six evaluated methods. Thus, the gain is not simply a more conservative
threshold; recalibration turns an informative but undercovered conditional
learner into a more usable book-level VaR forecast.

\subsection{Robustness Checks and Remaining Coverage Risks}
\label{sec:validation_threats}

The main validation risk is that the backtest may measure the wrong loss: selected contracts may miss the target moneyness, delta, or maturity; fallback marks or dropped observations may distort the sample; or date-$t$ features may leak the date-$t+1$ mark. Table~\ref{tab:validation} reports backtest diagnostics for the recalibrated VaR and three robustness specifications: strict marking uses only direct next-day contract marks, strict quality applies the target-quality screen, and no floor removes the nonnegative VaR floor.

\begin{table}[htp]
\caption{Backtest $p$-values (UC: Kupiec unconditional coverage; IND/CC: Christoffersen independence/conditional coverage) and robustness exceedance rates.}
\label{tab:validation}
\centering
\footnotesize
\setlength{\tabcolsep}{1.6pt}
\renewcommand{\arraystretch}{1.05}
\begin{tabular*}{\columnwidth}{@{\extracolsep{\fill}}llcccccc@{}}
\toprule
Market & Book
& $p_{\mathrm{UC}}$
& $p_{\mathrm{IND}}$
& $p_{\mathrm{CC}}$
& \makecell{Strict\\mark}
& \makecell{Strict\\quality}
& \makecell{No\\floor} \\
\midrule
SPX & \makecell[l]{ATM straddle}          & 0.219 & 0.850 & 0.462 & 0.107 & 0.107 & 0.111 \\
SPX & \makecell[l]{25d risk\\reversal}    & 0.872 & 0.047 & 0.138 & 0.100 & 0.108 & 0.105 \\
SPX & \makecell[l]{25d/10d\\put spread}   & 0.388 & 0.043 & 0.089 & 0.098 & 0.107 & 0.117 \\
QQQ & \makecell[l]{ATM straddle}          & 0.461 & 0.350 & 0.492 & 0.106 & 0.131 & 0.106 \\
QQQ & \makecell[l]{25d risk\\reversal}    & 0.336 & 0.212 & 0.288 & 0.101 & 0.106 & 0.108 \\
QQQ & \makecell[l]{25d/10d\\put spread}   & 0.584 & 0.106 & 0.233 & 0.100 & 0.103 & 0.106 \\
\bottomrule
\end{tabular*}
\end{table}

Strict marking, strict quality, and the no-floor check mostly preserve coverage
near the target. The main exceptions are QQQ ATM in the separate strict-quality run and the
SPX put spread without the floor. QQQ ATM has 616 dates passing the strict target-quality screen in the
main robust-marking backtest sample but only 374 forecasts after its rolling training period, with exceedance 0.131. Removing the floor raises
the SPX put spread exceedance to 0.117. For the SPX risk reversal and put spread, the independence test rejects at the
5\% level, but the unconditional-coverage and joint conditional-coverage tests
do not. This suggests marginal coverage is repaired, but residual exceedance clustering remains in asymmetric SPX books.

On dates retained by both strict and robust marking, normalized losses agree up
to numerical precision, so robust marking mainly expands the feasible panel.
Table~\ref{tab:learner_robustness} reports the learner robustness check.
GBRT and XGBoost remain close to the 10\% target, with book-level
recalibrated exceedance rates between 0.098 and 0.111.

\begin{table}[htp]
\caption{Reference learner robustness. GBRT denotes a standard
gradient-boosted regression-tree quantile learner. Book range is the min--max book-level recalibrated exceedance.}
\label{tab:learner_robustness}
\centering
\footnotesize
\setlength{\tabcolsep}{2.2pt}
\renewcommand{\arraystretch}{1.04}
\begin{tabular*}{\columnwidth}{@{\extracolsep{\fill}}llccc@{}}
\toprule
Market & Learner
& \makecell{Ref.\\exceedance}
& \makecell{Recal.\\exceedance}
& \makecell{Book\\range} \\
\midrule
SPX & LightGBM     & 0.171 & 0.106 & 0.099--0.111 \\
SPX & GBRT & 0.169 & 0.107 & 0.105--0.110 \\
SPX & XGBoost      & 0.155 & 0.107 & 0.105--0.108 \\
\midrule
QQQ & LightGBM     & 0.191 & 0.106 & 0.105--0.108 \\
QQQ & GBRT & 0.191 & 0.106 & 0.098--0.110 \\
QQQ & XGBoost      & 0.178 & 0.106 & 0.099--0.111 \\
\bottomrule
\end{tabular*}
\end{table}

\begin{figure*}[t]
\centering
\includegraphics[width=0.82\textwidth]{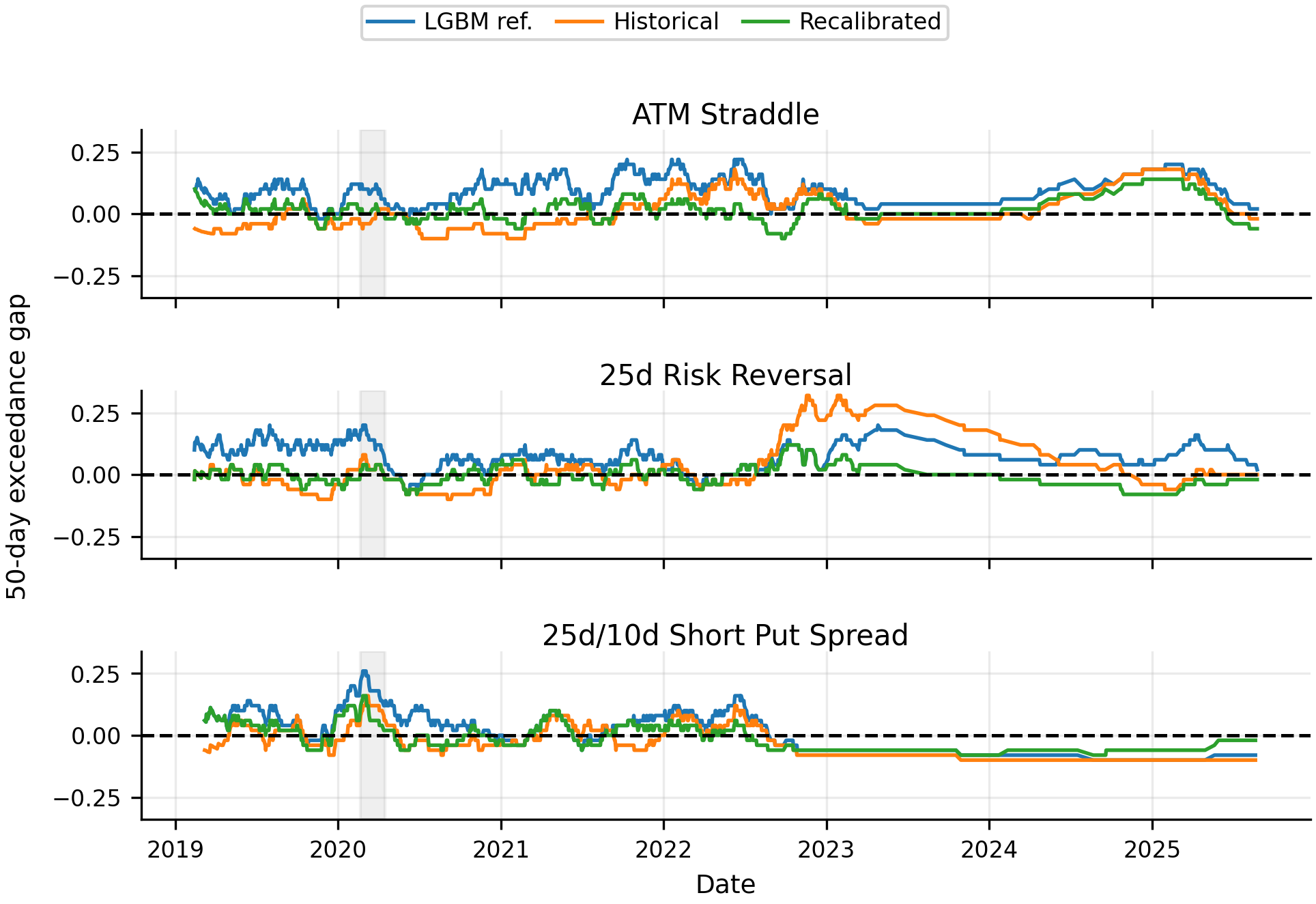}
\caption{SPX rolling 50-day exceedance gap relative to the 10\% target.
Positive = local undercoverage. Historical VaR shown as representative
classical baseline; shaded band marks the Feb--Apr 2020 stress window
(Section~\ref{subsec:dynamic}). Recalibration reduces persistent positive
gaps from the reference VaR; local clustering remains in asymmetric books.}
\label{fig:rolling_spx}
\end{figure*}

\subsection{Calibration Memory Ablation}

The main specification uses a 126-day residual window and exponential decay
parameter $\eta=0.01$. In Table~\ref{tab:calib_memory_ablation}, $W$ denotes
the residual window length and $\eta$ denotes the decay rate. The gap is
$100\max_b|\widehat p_b^{\mathrm{recal}}(W,\eta)-0.10|$ within each market,
where $\widehat p_b^{\mathrm{recal}}$ is the full-sample empirical exceedance
rate of the recalibrated VaR for book $b$. The max 50-day exceedance is the
largest rolling 50-day recalibrated exceedance rate across books. The ablation recomputes
only the residual calibration layer while holding the reference LightGBM
forecasts fixed.

\begin{table}[htp]
\caption{Calibration memory ablation. Gap is reported in percentage points (pp).}
\label{tab:calib_memory_ablation}
\centering
\footnotesize
\setlength{\tabcolsep}{3.0pt}
\begin{tabular*}{\columnwidth}{@{\extracolsep{\fill}}lcccc@{}}
\toprule
\makecell{Memory\\$(W,\eta)$}
& \makecell{SPX gap\\(pp)}
& \makecell{QQQ gap\\(pp)}
& \makecell{SPX max\\50-day\\exceedance}
& \makecell{QQQ max\\50-day\\exceedance} \\
\midrule
$(63,0.01)$        & 0.94 & 1.03 & 0.26 & 0.22 \\
$(126,0.00)$       & 1.23 & 0.82 & 0.28 & 0.24 \\
$(126,0.01)$ main  & 1.14 & 0.75 & 0.26 & 0.22 \\
$(126,0.03)$       & 1.15 & 1.09 & 0.22 & 0.20 \\
$(252,0.01)$       & 1.23 & 0.71 & 0.24 & 0.22 \\
\bottomrule
\end{tabular*}
\end{table}

Across all markets, books, and memory specifications, empirical exceedance rates remain close to the 10\% target, ranging from 0.0976 to 0.1123. The largest book-level coverage gap is 1.23 percentage points. The main $W=126,\eta=0.01$ setting is therefore not responsible for the
coverage repair; the improvement is stable across shorter windows, longer
windows, and alternative decay rates.

\subsection{Dynamic Diagnostics and Stress Periods}
\label{subsec:dynamic}

Figure~\ref{fig:rolling_spx} reports the SPX 50-day exceedance gap. The reference VaR has persistent positive exceedance gaps, especially around volatile periods. After recalibration, the maximum rolling 50-day exceedance falls from 0.32 to 0.24 for ATM, from 0.30 to 0.22 for the risk reversal, and from 0.36 to 0.26 for the put spread. QQQ shows the same direction of improvement, with recalibrated maximum 50-day exceedance no larger than 0.22 across the three books.

Crisis-window diagnostics show how the recalibrated VaR behaves during the February--April 2020 stress episode. In SPX, recalibrated crisis exceedance is 0.125 for the ATM straddle, 0.125 for the risk reversal, and 0.147 for the put spread. In QQQ, the corresponding rates are 0.077, 0.179, and 0.132. Because these windows contain roughly 32 to 39 dates, we use them as stress diagnostics rather than stand-alone validity tests. The largest stress-window miss is QQQ risk reversal, with crisis exceedance of 0.179. This suggests that the short residual window may adjust slowly when a book
with strong skew exposure faces a sharp move in volatility and skew.

\subsection{Target Probability Robustness}
\label{subsec:alpha_robustness}

Table~\ref{tab:alpha_robustness} reports the LightGBM reference and recalibrated VaR at $\alpha=0.05$ using the same rolling protocol as the main specification.

\begin{table}[htp]
\caption{Target probability robustness. Exceedance rates are pooled across
book-date observations within each market. Book range is the min--max
book-level recalibrated exceedance.}
\label{tab:alpha_robustness}
\centering
\footnotesize
\setlength{\tabcolsep}{3.0pt}
\renewcommand{\arraystretch}{1.04}
\begin{tabular*}{\columnwidth}{@{\extracolsep{\fill}}llccc@{}}
\toprule
$\alpha$ & Market
& \makecell{Ref.\\exceedance}
& \makecell{Recal.\\exceedance}
& \makecell{Book\\range} \\
\midrule
0.10 & SPX & 0.171 & 0.106 & 0.099--0.111 \\
0.10 & QQQ & 0.191 & 0.106 & 0.105--0.108 \\
0.05 & SPX & 0.111 & 0.057 & 0.055--0.059 \\
0.05 & QQQ & 0.124 & 0.057 & 0.054--0.059 \\
\bottomrule
\end{tabular*}
\end{table}

At $\alpha=0.05$, the reference VaR still undercovers, while residual
recalibration restores near-target coverage in both markets. Across the six book-market pairs, recalibrated exceedance ranges from 0.054 to
0.059, and the book-level unconditional-coverage tests do not reject at the
5\% level, with Kupiec $p$-values from 0.137 to 0.603.

\section{Conclusion and Future Work}
\label{sec:conclusion}

The empirical pattern in this paper points to two design principles
for option-book VaR. First, when the controlled position is defined by
multi-leg book construction and contract-level marking, the VaR target
is an operational object: book construction, next-day marking, loss
normalization, and the forecast-time information set are part of the
statistical target rather than implementation details outside it.
Second, once this target is fixed, conditional tail shape and marginal
level calibration are separable. A conditional learner can carry
useful tail signal while remaining miscalibrated in marginal coverage, and a
one-sided residual layer using only past forecast residuals repairs
the marginal level without refitting the conditional learner.
   
On the reported aggregate baseline metrics for SPX and QQQ, this decomposition delivers the lowest average violation, lowest pinball loss, and smallest maximum
rolling 50-day exceedance among the six compared methods. The recalibrated forecast passes unconditional coverage on all six book--market pairs and is broadly stable across the six robustness checks, including a stricter $\alpha=0.05$ target. Classical book-loss VaR baselines remain useful unconditional
benchmarks but do not match the severity--stability tradeoff at
comparable coverage. Residual exceedance clustering in SPX asymmetric
books remains visible, indicating that scalar level recalibration and
regime-aware dynamics are complementary risk control tasks.

Future work can extend the protocol in three directions. First,
regime-aware recalibration can adapt the residual window or decay rate during
stress periods, targeting local clustering in skew-sensitive books. Second,
marking uncertainty can be modeled directly instead of treating fallback marks
as deterministic. Third, large language models (LLMs) can be tested on dealer
commentary, option liquidity notes, macro news, and volatility commentary as
signals of marking stress~\cite{kirtac2024sentiment}, provided all inputs are
available at or before date $t$.

\noindent\textbf{Code and Artifact Availability.}
The code repository is available at
\href{https://github.com/TenghanZhong/option-book-var}{\nolinkurl{https://github.com/TenghanZhong/option-book-var}}.

\balance
\bibliographystyle{IEEEtran}
\bibliography{references}

\end{document}